\documentclass[showkeys,tightenlines,floatfix,noshowpacs,preprintnumbers,aps,superscriptaddress,pra,10pt,twocolumn,longbibliography,nofootinbib]{revtex4-2}

\usepackage{graphicx}
\usepackage{tikz}
\usepackage{tikz-cd}
\usetikzlibrary{decorations.markings,positioning}
\usetikzlibrary{shadows.blur}

\usepackage{physics}
\usepackage{float}  
\usepackage[english]{babel}
\usepackage[utf8]{inputenc}

\usepackage[colorlinks=true,linktocpage=true,breaklinks=true]{hyperref}
\hypersetup{
    allcolors  = {blue},
}
\usepackage{xurl} 

\usepackage{booktabs} 

\usepackage{amsmath}
\usepackage{amssymb} 
\usepackage{bbold}
\usepackage{amsthm}

\usepackage{xfrac}
\usepackage{mathtools}

\usepackage{standalone}

\usepackage{thmtools}
\usepackage{thm-restate}

\usepackage{listings}
\usepackage{enumitem}   

\usepackage[page,toc,titletoc,title]{appendix}

\usepackage{footmisc}
\usepackage{subcaption}

\usepackage{microtype} 
\usepackage[capitalize]{cleveref} 

\usepackage{nomencl}
\makenomenclature

\usepackage{soul}
\newtheorem{theorem}{Theorem}

\newtheorem{assump}{Assumption} 

\begin{document}

\title{An extended Wigner's friend no-go theorem inspired by generalized contextuality}

\author{Laurens Walleghem}\thanks{\href{laurens.walleghem@york.ac.uk}{laurens.walleghem@york.ac.uk}}
\affiliation{\small INL -- International Iberian Nanotechnology Laboratory, Braga, Portugal}
\affiliation{\small Department of Mathematics, University of York, York, United Kingdom}
\author{Lorenzo Catani}\thanks{\href{lorenzo.catani4@gmail.com}{lorenzo.catani4@gmail.com}}
\affiliation{\small INL -- International Iberian Nanotechnology Laboratory, Braga, Portugal}

\begin{abstract}
The renowned Local Friendliness no-go theorem demonstrates the incompatibility of quantum theory with the combined assumptions of Absoluteness of Observed Events -- the idea that observed outcomes are singular and objective -- and Local Agency -- the requirement that the only events correlated with a setting choice are in its future light cone. Granted that the Local Friendliness scenario can be realized, this result is stronger than Bell's theorem because the assumptions of Local Friendliness are weaker than those of Bell's theorem: Local Agency is less restrictive than local causality, and Absoluteness of Observed Events is encompassed within the notion of realism assumed in Bell's theorem. Drawing inspiration from the correspondence between nonlocality proofs in Bell scenarios and generalized contextuality proofs in prepare-and-measure scenarios, we present the Noncontextual Friendliness no-go theorem. This theorem demonstrates the inconsistency of quantum theory with the joint assumptions of Absoluteness of Observed Events and \textit{Noncontextual Agency}, the latter being a weaker version of noncontextuality, in the same way that Local Agency is a weaker version of local causality. Our result generalizes the Local Friendliness no-go theorem and, granted that the scenario can be realized, is stronger than no-go theorems based on generalized noncontextuality.
\end{abstract}

\maketitle

\section{Introduction} A promising approach to obtain a better understanding of quantum theory is to characterize its departure from the classical worldview. This can be rigorously done via no-go theorems, where one formally proves an inconsistency between the statistics of quantum theory and the statistics predicted under certain assumptions of what constitutes the classical worldview. In recent years, much attention has been paid to no-go theorems based on extended Wigner's friend experiments~\cite{brukner2017quantum,frauchiger2018quantum,bong2020strong,brukner2018no,cavalcanti2021implications,wiseman2023thoughtful,haddara2022possibilistic,nurgalieva2018inadequacy,vilasini2019multi,nurgalieva2020testing,Waaijer2021,allardGuerin2021nogotheorem,leegwater2022greenberger,ormrod2023theories,utreras2023allowing,walleghem2023extended,szangolies2020quantum,schmid2023review,walleghem2024connecting,nurgalieva2023multi,walleghem2024strong,montanhano2024wigner,walleghem2026freechoicesabsoluteinternalized,walleghem2026wignersfriendsblackhole}. The latter highlight the interpretational issues that arise when involving agents described as quantum systems that perform measurements.

A prominent example of such no-go theorems is the Local Friendliness (LF) no-go theorem~\cite{bong2020strong,cavalcanti2021implications,haddara2022possibilistic}, which shows a contradiction between the predictions of quantum theory and the assumptions of Absoluteness of Observed Events -- ascribing single and objective values to observed measurement outcomes  -- and Local Agency -- a strengthened version of no-signalling that also applies to situations involving outcome correlations that cannot be verified by a single agent.
The LF no-go theorem can be conceptualized as combining a Wigner's friend experiment with a Bell scenario and, assuming that the LF scenario can be realized, results in an even stronger theorem than Bell's theorem, given that the LF assumptions are weaker than the ones of Bell's theorem~\cite{wiseman2023thoughtful}. One way to ensure that the LF scenario can be realized is to assume a form of universality of quantum theory, which allows for the existence of observers that can be described quantum mechanically, and of an advanced agent who can, in principle, apply an arbitrary unitary transformation to them. The sense in which we claim that the LF no-go theorem is said to be stronger than Bell's theorem is exemplified in \cite{haddara2024local}, where the authors show that, in standard LF scenarios, the polytope of correlations consistent with Local Friendliness (the conjunction of Absoluteness of Observed  Events and Local Agency) is strictly larger than the polytope of locally causal correlations.

In this work, we combine Wigner's friend experiments with scenarios manifesting generalized contextuality~\cite{spekkens2005contextuality}, one of the leading notions of nonclassicality. We consider the simplest scenario showing generalized contextuality~\cite{pusey2018robust}, which is a prepare-and-measure scenario on a single system involving four preparations and two measurements, and we include quantum observers. In designing the experiment we leverage a known correspondence between proofs of nonlocality in Bell scenarios and proofs of contextuality in prepare-and-measure scenarios~\cite{liang2010specker,CoeckeKissinger2017,Henaut2018,CataniFaleiro2023, Wright_2023}. In the same way in which, under the assumption that the scenario can be realized, the LF no-go theorem  is stronger than Bell's theorem, we obtain a stronger no-go theorem than the one associated with generalized contextuality.

The remainder of the manuscript is organized as follows. In \Cref{sec:background}, we review the necessary background on extended Wigner's friend experiments, Bell nonlocality, and generalized contextuality. 
\Cref{sec:LF} presents the Local Friendliness no-go theorem, while \Cref{sec:NF} introduces the Noncontextual Friendliness no-go theorem. 
We conclude in \Cref{sec:discussion} with a discussion of our findings.

\section{Background} \label{sec:background}

\subsection{Extended Wigner's friend experiments}
The notorious measurement problem~\cite{Maudlin1995} arises because quantum theory prescribes two distinct types of time evolution for the state of a quantum system: (i) unitary, deterministic evolution for the state of a closed quantum system, and (ii) an indeterministic evolution for the state of the system after it undergoes a measurement. Wigner's famous thought experiment~\cite{wigner1995remarks} illustrates this tension as follows.
Wigner's friend performs a measurement in a sealed lab. Wigner, as her \textit{superobserver}, ascribes a unitary evolution (i) to her lab, while the friend's perspective follows the indeterministic evolution (ii).

In recent years, no-go theorems based on extensions of Wigner's friend argument have been developed.
These shift the primary tension from being between unitarity and the measurement postulate to a subtler one between unitarity and the Born rule~\cite{schmid2023review}. Specifically, these theorems leverage the fact that unitary operations allow Wigner to undo the friend's measurement (erasing her outcomes) while simultaneously requiring that one obtains a joint probability distribution for the observed outcomes -- including those erased -- according to the Born rule.

The most prominent extended Wigner's friend no-go theorems combine Wigner's original setup with Bell's scenario. The first attempt was made by Brukner~\cite{brukner2015,brukner2018no}, who aimed to bypass the counterfactual nature of Bell's theorem by ensuring that each outcome in the scenario could actually be observed by some agent.
Indeed, Bell's theorem appeals to hidden variables to assign definite outcomes even to measurements that are not performed. However, critics such as Asher Peres argued against this counterfactual aspect, famously stating that ``unperformed experiments have no results'' ~\cite{peres1978unperformed}. Brukner's insight was that by integrating Bell's setup with Wigner's, one could construct a scenario involving only observed events, with no counterfactual outcomes. Although Brukner's no-go theorem fell short of fully achieving this goal, the Local Friendliness no-go theorem~\cite{bong2020strong} succeeded.

\subsection{Bell nonlocality and generalized contextuality}
In a Bell scenario, two parties, Alice and Bob, share a bipartite system and each perform a measurement on their respective parts.
Alice's and Bob's measurement choices are denoted by $x$ and $y$, respectively, with outcomes $a$ and $b$.
Repeating the protocol many times allows for the collection of measurement statistics $\wp(a,b|x,y)$.\footnote{We here introduce, similarly to \cite{bong2020strong}, a notational difference between empirical probability distributions that can be obtained by a single agent from the collected data, denoted with $\wp$, and theoretical probability distributions required to exist given certain assumptions, but potentially not observable by any agent, denoted with $p$.} These correlations are consistent with the existence of an ontological model satisfying \textit{local causality}~\cite{bell1964einstein} if they can be expressed as follows:
\begin{equation}
\label{eq:locality}
    \wp(a,b|x,y) = \sum_\lambda p(\lambda) p(a|x,\lambda)p(b|y,\lambda),
\end{equation}
where $\lambda$ represents the ontic state in the ontological model.
The \textit{ontological models framework}~\cite{harrigan2010einstein} formally provides a realist explanation of the statistics of a physical theory by postulating the existence of physical systems with properties (ontic states) represented by points $\lambda$ in a measurable set $\Lambda$ (the ontic state space). The statistics predicted by the theory are reproduced, in the ontological model, via the law of classical total probability: $\wp(k|M,P)= \sum_\lambda p(k|M,\lambda)p(\lambda|P),$ where the preparation $P$ is represented by a probability distribution $p(\lambda|P)$, and the measurement $M$ with outcome $k$ is represented by a probability distribution $p(k|M,\lambda)$.

An equivalent way to require that Eq.~\eqref{eq:locality} is satisfied is to demand the existence of a joint distribution
$p(a_0,a_1,b_0,b_1)$, where $a_x=a|x, b_y = b|y$,
for all possible measurement outcomes and settings, from which the observed probabilities $\wp(a,b|x,y)$ can be obtained by marginalizing over the unperformed measurements~\cite{fine1982joint,abramsky2011sheaf}, for example, $\wp(a,b|x=0,y=0) = \sum_{a_1,b_1}p(a_0,a_1,b_0,b_1)$.
Bell's theorem famously proves that quantum theory manifests \textit{nonlocality}, \textit{i.e.}, that it does not admit of a locally causal ontological model~\cite{bell1964einstein}.

\begin{figure}
         \centering \includegraphics[width=0.45\textwidth,height=0.13\textheight]{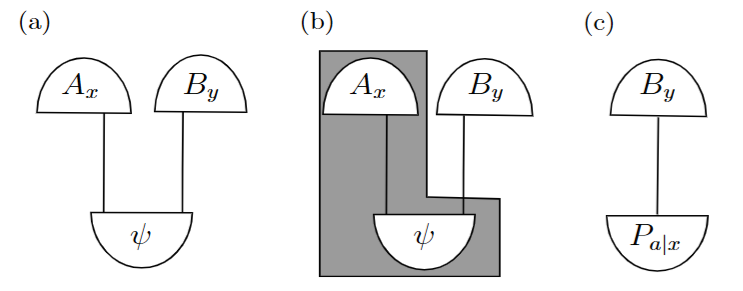}
         \caption{(a) Bell scenario; (b) Bell scenario as a prepare-and-measure scenario; (c) prepare-and-measure scenario. Subscripts denote the preparation and measurement variables, and $\psi$ the bipartite state.}
         \label{fig:Bell_to_PM}
\end{figure}

A Bell scenario can also be conceptualized as a prepare-and-measure scenario, where the preparation stage in Bob's side is determined by steering from Alice's setting and outcome $(x,a)$, and the measurement stage consists of Bob's measurement $y$ and outcome $b$ \cite{CoeckeKissinger2017,Henaut2018,CataniFaleiro2023,Wright_2023} (see \Cref{fig:Bell_to_PM}). In this setting, the nonlocality of quantum theory turns into the inconsistency of quantum theory with a generalized noncontextual ontological model~\cite{spekkens2005contextuality}.

Generalized noncontextuality requires that operational equivalences predicted by a physical theory correspond to identities in the ontological model of the theory. For instance, if two preparations $P$ and $P'$ are operationally equivalent -- \textit{i.e.}, for all possible measurements $M$ with outcomes $k$, $\wp(k|M,P) = \wp(k|M,P')$ -- a preparation noncontextual ontological model represents them with identical probability distributions: $p(\lambda|P) = p(\lambda|P')$. A similar definition applies to measurements.

The credentials for generalized noncontextuality lie in a methodological principle inspired by Leibniz's principle of the identity of indiscernibles \cite{Spekkens2019Leibniz}, or, relatedly, by the principle of no operational fine-tuning~\cite{catani2023mathematical}. Generalized noncontextuality satisfies the desiderata for a good notion of classicality, as argued, for example, in the introduction of \cite{CataniGalleyGonda2024} and in \cite{SchmidSolstice2022}. In the context of local causality, no-signalling is the operational equivalence that is preserved in the ontological model. Specifically, with respect to $a$ in a Bell scenario,  no-signalling implies $\wp(a|x,y)=\wp(a|x)$ and its preservation in the ontological model reads as $p(a|\lambda,x,y) = p(a|\lambda,x).$
This assumption, known as parameter independence, combined with outcome independence -- \textit{e.g.}, with respect to $a$,   $p(a|\lambda,b,x,y)=p(a|\lambda,x,y)$ -- defines local causality\footnote{Note that the assumption of measurement independence, $p(\lambda|x,y)=p(\lambda)$, which is also required to prove Bell's theorem, is already built into the definition of ontological models~\cite{catani2023mathematical}.}.

\begin{figure}[h]
         \centering \includegraphics[width=0.29\textwidth]{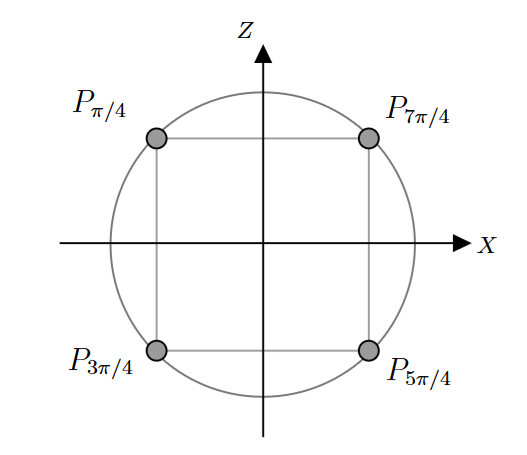}
         \caption{The simplest nontrivial scenario~\cite{pusey2018robust},
         involving four preparations (gray dots) and two measurements (the two axes) that provide a maximal violation of the noncontextuality inequalities.}
         \label{fig:Bloch}
\end{figure}

The simplest proof that quantum theory manifests \textit{contextuality}, \textit{i.e.}, it does not admit of a generalized noncontextual ontological model, can be constructed in the \textit{simplest nontrivial scenario}~\cite{pusey2018robust,pusey2019,Khoshbin_2024}. This is a prepare-and-measure scenario consisting of two tomographically complete measurements, \textit{e.g.}, the Pauli $X$ and $Z$ measurements on a qubit,
 and four preparations, \textit{e.g.}, $P_{\theta}= \cos(\theta/2)\ket{0}+\sin(\theta/2)\ket{1}$, for $\theta = \pi/4,3\pi/4,5\pi/4,7\pi/4$. These are represented on the Bloch disk in ~\Cref{fig:Bloch}.
The preparations satisfy the following operational equivalence, \begin{align}
    &\frac{1}{2}\wp(k|M,P_\frac{\pi}{4}) +  \frac{1}{2}\wp(k|M,P_\frac{5\pi}{4}) \nonumber
    \\ &=  \frac{1}{2}\wp(k|M,P_\frac{3\pi}{4}) +   \frac{1}{2}\wp(k|M,P_\frac{7\pi}{4}).
\end{align}
However, there does not exist an ontological model where these also provide an ontological equivalence.

A strategy for obtaining a maximum quantum violation in a Bell scenario can be mapped to the simplest scenario with the specific choices of preparations and measurements above. In the former case, one shows nonlocality; in the latter, preparation contextuality.

\section{Local Friendliness no-go theorem~\cite{bong2020strong}} \label{sec:LF}
\begin{figure}[h]
         \centering \includegraphics[width=0.45\textwidth,height=0.142\textheight]{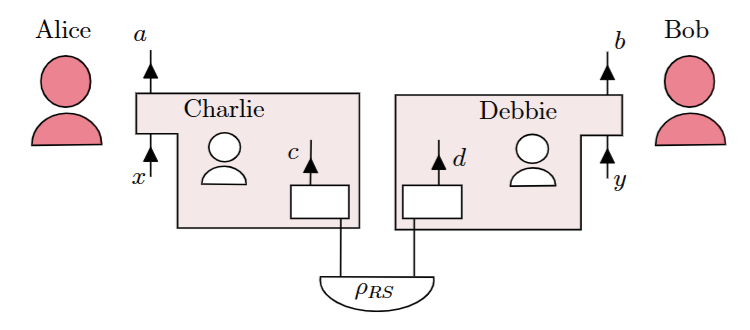}
         \caption{Local Friendliness setup~\cite{schmid2023review}.
         }
         \label{fig:LF_cartoon}
\end{figure}

The experimental setup, illustrated in \Cref{fig:LF_cartoon}, is described as follows. Charlie and Debbie, spacelike separated, share a bipartite system in the state $\rho_{RS}$.
They perform measurements denoted with $C,D$ on their respective subsystems, $R,S$, obtaining outcomes $c,d\in\{0,1\}$. The measurements are modelled by unitaries $U_C,U_D$, respectively.
After Charlie has performed his measurement, depending on the value of the variable $x\in\{0,1\}$, Alice makes one of two choices.
If $x=0$ she takes Charlie's outcome to be her outcome.\footnote{Here, an additional ``tracking assumption'' is introduced. For a discussion on why this is considered a minimal assumption, see~\cite{schmid2023review}.} If $x=1$ she acts as a superobserver and undoes Charlie's measurement by applying $U_C^\dagger$, thereby erasing the outcome record of $c$; subsequently she performs a measurement denoted with $A$ on $R$, obtaining outcome $a\in\{0,1\}$.
Similarly, Bob makes one of two choices associated with $y\in\{0,1\}$. For $y=0$, he takes Debbie's outcome to be his outcome. For $y=1$ he undoes Debbie's measurement and subsequently performs a measurement denoted with $B$ on $S$, obtaining outcome $b\in\{0,1\}$.

Given the setup, let us state the following assumptions.~\begin{assump}[Absoluteness of Observed Events (AOE)] \label{assumption:AOE}
An observed event is an absolute single event, not relative to anything or anyone.
\end{assump}
This assumption can be motivated as an instance of realism and, in the LF scenario, it implies that there exist distributions $p(a,b,c,d|x,y)$ from which one can obtain the empirical probabilities, $\wp(a,b|x=1,y=1),\wp(a,d|x=1,y=0),\wp(c,b|x=0,y=1),\wp(c,d|x=0,y=0)$ via marginalisation.

\begin{assump}[Local Agency] \label{ass:LA}
Any instance of no-signalling that is verifiable when all relevant variables are made jointly accessible still holds when those variables are not jointly accessible to any single agent.
\end{assump}

The above formulation makes explicit the strong reading of the standard Local Agency assumption~\cite{cavalcanti2021implications} (``The only relevant events correlated with an intervention are in its future light cone''), in which ``relevant events'' include the jointly inaccessible ones whose existence is guaranteed by AOE.
In the LF scenario, Local Agency implies, for example, that $p(c,b|x,y=1) = p(c,b|y=1)$, as none of $c,b$ are obtained in the future of the choice $x$.
Notice how this is different from requiring the disjoint conditions to hold: $\wp(b|x,y=1)=\wp(b|y=1)$ -- an instance of no-signalling -- and $\wp(c|x,y)=\wp(c)$ -- an instance of no-signalling to the past, or no-superdeterminism.
Local Agency is stronger than standard no-signalling and no-superdeterminism insofar as it requires the joint independence, which, in addition to the no-signalling and no-superdeterminism above, also requires $p(b|c,x,y=1)=p(b|c,y=1)$.

Local Agency can be motivated by a commitment to relativistic principles and the belief that these should hold even if they cannot be directly verified by a single agent. It can also be justified as an instance of no operational fine-tuning~\cite{catani2023mathematical} in a universe where AOE holds: if events are absolute and a theory prescribes no-signalling, the latter should also hold even if it cannot be verified by a single agent. It would otherwise be conspiratorial for the prescription to fail precisely on those analogues that, while guaranteed to exist by AOE, are not verifiable by a single agent.

It is worth noting that while Local Agency is stronger than no-signalling, it is weaker than local causality, as shown in \cite{cavalcanti2021implications}.
This is further evident in scenarios where the polytope of correlations implied by Local Friendliness (the conjunction of AOE and Local Agency) is strictly larger than the polytope of locally causal correlations, as demonstrated in \cite{haddara2024local}.

We can now state the LF no-go theorem~\cite{bong2020strong,haddara2022possibilistic,cavalcanti2021implications}.
\begin{theorem}[LF no-go theorem] \label{th:LF}
If a superobserver can perform arbitrary quantum operations on an observer and its environment, then no physical theory can satisfy Local Friendliness.
\end{theorem}
\noindent The proof of the theorem is contained in Appendix A. We also emphasize that, in terms of rigor, the LF no-go theorem has the same status as Bell's theorem, since it is precise, theory-independent, and -- at least in principle -- empirically testable~\cite{wiseman2023thoughtful}.

\section{Noncontextual Friendliness no-go theorem} \label{sec:NF}
\begin{figure}[h!]
         \centering \includegraphics[width=0.26\textwidth,height=0.22\textheight]{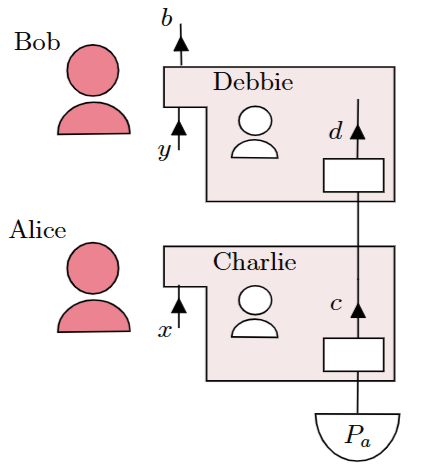}
         \caption{Noncontextual Friendliness setup.}
         \label{fig:OF_tryout}
\end{figure}

The experimental setup, illustrated in \Cref{fig:OF_tryout}, is described as follows.
Alice prepares a qubit $S$ in the state $P_a$, for values of $a=0,1$ each happening with probability $1/2$.
She sends the prepared system to her friend Charlie, who performs a bipartite measurement $C$ on $S$, where the two possible outcomes are labelled by $0$ and $1$. The measurement is modelled unitarily as $U_C$.
After Charlie has performed his measurement, depending on the value of the variable $x\in\{0,1\}$, Alice makes one of two choices.
If $x=0$ she takes Charlie's outcome to be her outcome. If $x=1$ she acts as a superobserver and undoes Charlie's measurement by applying $U_C^\dagger$, thereby erasing the outcome record of $c$. Notice that, in this latter case, unlike the LF setup, she does not perform a measurement $A$ on $R$ after erasing the outcome of Charlie.
Subsequently, the system is passed to Debbie who performs a measurement $D$ on the system. This measurement is modelled unitarily as $U_D$.
After Debbie has performed her measurement, depending on the value of the variable $y\in\{0,1\}$, Bob makes one of two choices.
If $y=0$ he takes Debbie's outcome to be his outcome. If $y=1$ he acts as a superobserver and undoes Debbie's measurement by applying $U_D^\dagger$, thereby erasing the outcome record of $d$, and subsequently performs a measurement denoted with $B$ on $S$ yielding outcome $b$.

Given this setup, let us state two assumptions:
 AOE (Assumption 1) and Noncontextual Agency (Assumption 3), whose conjunction we call \textit{Noncontextual Friendliness} (NF).

\begin{assump}
[Noncontextual Agency]
\label{Ass:OA}
Any operational equivalence that is verifiable when all relevant variables are made jointly accessible still holds when those variables are not jointly accessible to any single agent.
\end{assump}

Notice the similarity between Assumption \ref{ass:LA} and Assumption \ref{Ass:OA}: the former can be viewed as a special case of the latter, where the operational equivalence considered is specifically no-signalling. Just as Local Agency is a stronger version of no-signalling, Noncontextual Agency is a stronger version of standard operational equivalence. In the context of the setup considered here, standard operational equivalence pertains to setting choices. More precisely, they are about the preparations associated with choices $x=0$ and $x=1$: $\wp(d|x,y=0)=\wp(d|y=0)$ and $\wp(b|x,y=1)=\wp(b|y=1)$.
Noncontextual Agency additionally implies that these equalities also hold when conditioned on $c$, \textit{i.e.}, $p(d|c,x,y=0)=p(d|c,y=0)$ and $p(b|c,x,y=1)=p(b|c,y=1)$, despite the fact that no single observer can verify them, since when $x=1$, the outcome $c$ is erased. Thus, similarly to Local Agency, Noncontextual Agency lifts the disjoint conditional independences of $d|y$ and $c|y$ on $x$ to a joint conditional independence of $c,d|y$ on $x$ (and similarly for $b,c|x,y$).
Regarding the name ``Noncontextual Agency'', the term noncontextual is used similarly to generalized noncontextuality: just as generalized noncontextuality requires that operational equivalences are preserved as ontological identities within an ontological model explaining the physical theory, here noncontextual indicates that operational equivalences are preserved at the level of the Absoluteness of Observed Events -- and thus remain valid even when they are operationally inaccessible, in the sense that no single agent can verify them. In other words, it represents a form of noncontextuality formulated not within the framework of ontological models~\cite{harrigan2010einstein}, but under the sole assumption that the underlying reality satisfies AOE. This terminology parallels that of Local Agency, where local is inspired by the notion of local causality in Bell's theorem (specifically, its parameter independence aspect, sometimes referred to as locality~\cite{wiseman2017causarum}), but where no-signalling is required to hold only at the AOE level -- \textit{i.e.}, in a world where observed events are absolute -- rather than in an ontological model.

Formally, a physical theory specifies, for any pair of operationally indistinguishable preparations $P, P'$, that the equality $\wp(k|M, P) = \wp(k|M, P')$ holds for every outcome $k$ and measurement $M$ in principle allowed by the theory. Under AOE, the correlations among jointly inaccessible outcomes are events of the same ontological kind as accessible ones -- they exist as facts about the world even when no single agent can verify them. Noncontextual Agency is the requirement that the theory's predicted operational equivalences extend to these jointly inaccessible events. The justification is the methodological principle of no fine-tuning~\cite{Spekkens2019Leibniz,catani2023mathematical}: if the theory predicts the equivalence to hold whenever it is operationally accessible, then -- absent some structural reason that distinguishes accessible from inaccessible cases -- it would be conspiratorial for the equivalence to fail precisely (and only) on the inaccessible cases. Drawing such a distinction would require the theory to treat empirical verifiability itself as a privileged structural feature, in tension with realism about observed events as encoded by AOE.

In the present setup, the relevant operational equivalences are as follows. First, the theory predicts the marginal independence $\wp(c|x,y)=\wp(c)$, since $c$ lies in the past of $x,y$ (this independence is also derivable without reference to any external observer, as shown in Section 4.2.2 of Ref.~\cite{ying2023relating}). Second, the theory predicts the conditional equivalence $\wp(d|c,x,y)=\wp(d|c,y)$: in any alternative protocol where $c$ is recorded in a separate register prior to Alice's operation, the system $S$ is in the post-measurement state $P_c$ regardless of whether Alice subsequently applies $U_C^\dagger$ -- the undoing acts on $S$ but not on the external record -- and Debbie's measurement on $S$ then yields a distribution independent of $x$. An analogous argument establishes $\wp(b|c,x,y)=\wp(b|c,y)$. Noncontextual Agency promotes these in-principle predictions to constraints on the empirical-level correlations in the actual protocol, where $c$ is erased and the joint variables $(c,d)$ or $(c,b)$ are inaccessible to any single agent but, by virtue of AOE, still exist as facts about the world.

It is worth noting that while Noncontextual Agency is stronger than operational equivalence, it is weaker than generalized noncontextuality: in Appendix B, we present a scenario showing that the polytope of NF correlations is strictly larger than the polytope of noncontextual correlations, mirroring the proof strategy of \cite{haddara2024local}.

We can now state the NF no-go theorem.
\begin{theorem}[Noncontextual Friendliness no-go theorem] \label{th:NCF}
If a superobserver can perform arbitrary quantum operations on an observer and its environment, then no physical theory can satisfy Noncontextual Friendliness.
\end{theorem}

\proof
Let us consider the NF scenario of \Cref{fig:OF_tryout} with the following specifications (the preparations and measurements employed will be the same as in the simplest nontrivial scenario). Alice chooses preparations  $P_{0}= P_{\pi/4}$ and $P_1=P_{5\pi/4}$, for $a=0,1$, respectively.
Charlie measures in the basis given by $P_{3\pi/4}$ and $P_{7\pi/4}$.
Debbie's measurement $D$ is the Pauli $Z$ measurement.
When $y=1$, Bob undoes Debbie's measurement and performs a Pauli $X$ measurement on $S$.
At the end of the experiment the following empirical correlations of observed outcomes are obtained, \begin{equation} \label{eq:empirical_prob_LF}
    \begin{split}
        &\wp(c,d \mid x=0, y=0), \\
&\wp(c,b\mid x=0, y=1), \\
&\wp(a,d \mid x=1, y=0), \\
&\wp(a,b\mid x=1, y=1).
    \end{split}
\end{equation}

By AOE and Noncontextual Agency, we can rewrite these empirical correlations as follows,

\begin{align} \label{eq:empirical_PandM_EWF_1}
        &\wp(c,d \mid x=0, y=0) = p(c,d|x=1,y=1), \\ \label{eq:empirical_PandM_EWF_2}
&\wp(c,b\mid x=0, y=1)  = p(c,b|x=1,y=1), \\ \label{eq:empirical_PandM_EWF_3}
&\wp(a,d \mid x=1, y=0)  = p(a,d|x=1,y=1), \\ \label{eq:empirical_PandM_EWF_4}
&\wp(a,b\mid x=1, y=1) = p(a,b|x=1,y=1).
\end{align}

AOE is used to consider joint probability distributions when $c$ or $d$ are involved and possibly erased.
To obtain Eq.~\eqref{eq:empirical_PandM_EWF_1}, we first appeal to an instance of Noncontextual Agency that enforces no-superdeterminism, $p(c,d|x,y){=}p(c,d|x)$, yielding $p(c,d|x{=}0,y{=}0){=}p(c,d|x{=}0,y{=}1)$; we can then apply the rule of conditional probability and consider two other instances of Noncontextual Agency (the first again enforcing no-superdeterminism): $p(c|x,y){=}p(c)$ and $p(d|c,x,y){=}p(d|c,y)$, thus obtaining \[\begin{split}&\wp(c,d|x{=}0,y{=}1) = p(c|x{=}0,y{=}1)p(d|c,x{=}0,y{=}1) \\&=p(c|x{=}1,y{=}1)p(d|c,x{=}1,y{=}1)=p(c,d|x{=}1,y{=}1).\end{split}\]

Similar instances of Noncontextual Agency, but with  variables $c,b$ -- namely $p(c|x,y)=p(c)$ and $p(b|c,x,y)=p(b|c,y)$ --  lead to Eq.~\eqref{eq:empirical_PandM_EWF_2},
\[\begin{split}&\wp(c,b|x{=}0,y{=}1) = p(c|x{=}0,y{=}1)p(b|c,x{=}0,y{=}1) \\ &=p(c|x{=}1,y{=}1)p(b|c,x{=}1,y{=}1)=p(c,b|x{=}1,y{=}1).
\end{split}\]

Eq.~\eqref{eq:empirical_PandM_EWF_3} derives from an instance of Noncontextual Agency enforcing no-superdeterminism, $p(a,d|x,y)=p(a,d|x)$.

By assuming AOE, we can assert that all the correlations in Eqs.~\eqref{eq:empirical_PandM_EWF_1}, \eqref{eq:empirical_PandM_EWF_2}, \eqref{eq:empirical_PandM_EWF_3} and \eqref{eq:empirical_PandM_EWF_4} can be obtained from a single joint probability distribution $p(a,b,c,d|x=1,y=1)$ via marginalization.
However, recall that the preparations and measurements used to obtain these correlations correspond to those in the simplest scenario that achieve the maximum violation of noncontextuality inequalities~\cite{pusey2018robust}. Through the mapping to the Bell scenario, they are also equivalent to those achieving the maximum violation of Bell inequalities. By Fine's theorem~\cite{fine1982hidden,abramsky2011sheaf}, such a joint probability distribution cannot exist.
Let us stress that the NF assumptions are weaker than those used in the noncontextuality no-go theorem proven in the simplest scenario~\cite{pusey2018robust}. Only the empirical correlations in the scenarios are the same.
Notice how the present proof mirrors the proof of the LF no-go theorem contained in Appendix A, with Noncontextual Agency replacing Local Agency.

\section{Discussion} \label{sec:discussion}
In this work, we have shown that the Local Friendliness no-go theorem---arguably the strongest existing no-go theorem highlighting the tension between quantum theory and the classical worldview---can be generalized to the Noncontextual Friendliness no-go theorem. Just as no-signalling is a specific instance of an operational equivalence, Local Agency is a specific instance of Noncontextual Agency.
Furthermore, just as the LF no-go theorem is stronger than Bell's theorem, granted that the LF scenario can be realized, the NF no-go theorem is stronger than theorems based on generalized noncontextuality, particularly the one leveraging the simplest scenario, granted that the NF scenario can be realized.
Local Agency strengthens no-signalling by requiring it to hold even when it cannot be verified by a single agent. Similarly, Noncontextual Agency extends operational equivalence to cases where no single agent can verify the equivalence.

Key to establishing these connections is the rephrasing of Local Agency, highlighting how it can be understood as a stronger version of no-signalling. Specifically, Local Agency requires that the no-signalling predicted by the theory hold not only for operationally accessible correlations but also for the jointly inaccessible analogues that exist by virtue of AOE. Recognizing that no-signalling is itself an instance of operational equivalence naturally leads to the mirrored definition of Noncontextual Agency.

It is worth discussing here the conceptual significance of our result, which primarily arises from the conceptual status of Noncontextual Agency. In our formulation, the assumption of Noncontextual Agency constitutes an instance of the methodological principle of no fine-tuning: in a world where AOE holds, it would be conspiratorial for operational equivalences predicted by the theory to hold whenever they are operationally accessible but to fail precisely on the jointly inaccessible cases. It is indeed natural (i.e., non-conspiratorial) to assume that, in a world satisfying AOE, the theory continues to predict equivalences to hold for correlations involving jointly inaccessible outcomes.

Notice that the above considerations also apply to the case of Local Agency, which is an instance of Noncontextual Agency. Consequently, Local Agency shifts from being justified by a metaphysical assumption about the nature of the underlying causal structure---namely, that it must be relativistic---to being an instance of the methodological principle of no fine-tuning. As such, it carries a lighter ontological weight.
In turn, the lesson drawn from the LF and NF no-go theorems becomes clear: if one believes that the LF and NF scenarios can indeed be realized -- for instance, by endorsing the universality of quantum theory\footnote{In support of the universality of quantum theory, we note that there is currently no experimental evidence against it, nor any indication of regimes in which it might fail. Moreover, even if quantum theory were not universal, one would still be required to specify precisely the domains of its applicability and to provide a clear account of how the transition to alternative physical descriptions occurs in other regimes.} -- and accepts the no fine-tuning (or no-conspiracy) principle -- which, following many others \cite{Spekkens2019Leibniz,catani2023mathematical,schmid2021unscrambling}, we regard as methodologically indispensable -- then one is compelled to conclude that observed events are relative; that is, one must abandon AOE, the sole remaining metaphysical assumption in the theorem.

Interpreting Local Agency as an instance of Noncontextual Agency also allows us to dig deeper into the justification of Local Agency itself. One route grounds it in a fundamental relativistic causal structure taken to hold also for jointly inaccessible outcomes -- a second metaphysical commitment beyond AOE. A second route, the one this work pursues, motivates it as an instance of the no fine-tuning principle in a world where AOE holds -- a methodological rather than metaphysical stance. Both routes are available; for those who regard ontological parsimony as a virtue, the second is the more natural companion to AOE. Why should one believe that no-signalling also holds for jointly inaccessible outcomes, and that the relativistic causal structure extends to them? Because, in a world where AOE holds, it would otherwise be conspiratorial not to.

As far as we are aware, the only work that explicitly discusses no fine-tuning in the context of an extended Wigner's friend scenario is \cite{ying2023relating}, although the authors do so within the framework of causal models. They show that, even when one allows for a nonclassical causal model and assumes AOE, it remains impossible to account for the statistics of the LF scenario without invoking causal fine-tuning~\cite{Wood_2015,Cavalcanti_2018} -- a result which bears a close relation to ours. We leave a detailed investigation of this connection for future work.

Another interesting avenue for future research would be to explore proofs of the NF no-go theorem in scenarios beyond the one considered here, which is based on the simplest scenario.
More broadly, it would be valuable to generalize the NF no-go theorem by considering examples of fine-tunings that go beyond those linked to contextuality~\cite{catani2023mathematical}, such as violations of bounded ontological distinctness~\cite{Chaturvedi_2020,Khoshbin_2024}. Additionally, it would be worthwhile to investigate NF scenarios that are realizable within post-quantum theories -- for example those proposed in~\cite{vilasini2019multi,ormrod2023theories,nurgalieva2025theoryadmitswignersfriend}.
It should also be noted that other extended Wigner's friend no-go theorems based on \textit{Kochen-Specker contextuality}~\cite{kochen1990problem} exist~\cite{walleghem2023extended,szangolies2020quantum}.
In this context, our assumption of Noncontextual Agency encompasses the newly introduced assumption of Commutation Irrelevance adopted therein.

\section*{Acknowledgements}
The authors thank David Schmid, Rafael Wagner and Yìlè Y{\=\i}ng for helpful discussions, particularly those that prompted a deeper understanding of the motivations behind Noncontextual Agency and its relation to Local Agency. LW thanks Rui Soares Barbosa, Eric Cavalcanti and Raman Choudhary for insightful discussions.
LW acknowledges support from the United Kingdom Engineering and Physical Sciences Research Council (EPSRC) DTP Studentship (grant number EP/W524657/1). LW also thanks the International Iberian Nanotechnology Laboratory--INL in Braga, Portugal and the Quantum and Linear-Optical Computation (QLOC) group for the kind hospitality. LC acknowledges funding from the Horizon Europe
project FoQaCiA, GA no.101070558.

\bibliography{refs}

@article{frauchiger2018quantum,
  title={Quantum theory cannot consistently describe the use of itself},
  author={Frauchiger, Daniela and Renner, Renato},
  journal={Nat. Commun.},
  volume={9},
  number={1},
  pages={3711},
  year={2018},
  publisher={Nature Publishing Group UK London},
  doi={https://doi.org/10.1038/s41467-018-05739-8}
}

@article{ying2023relating,
  doi = {10.22331/q-2024-09-26-1485},
  url = {https://doi.org/10.22331/q-2024-09-26-1485},
  title = {Relating {W}igner's {F}riend {S}cenarios to {N}onclassical {C}ausal {C}ompatibility, {M}onogamy {R}elations, and {F}ine {T}uning},
  author = {Yīng, Y{\`{i}}l{\`{e}} and Ansanelli, Marina Maciel and Biagio, Andrea Di and Wolfe, Elie and Schmid, David and Cavalcanti, Eric Gama},
  journal = {{Quantum}},
  issn = {2521-327X},
  publisher = {{Verein zur F{\"{o}}rderung des Open Access Publizierens in den Quantenwissenschaften}},
  volume = {8},
  pages = {1485},
  month = sep,
  year = {2024}
}

@misc{nurgalieva2025theoryadmitswignersfriend,
      title={Any theory that admits a {W}igner's {F}riend type multi-agent paradox is logically contextual}, 
      author={Nuriya Nurgalieva and V. Vilasini},
      year={2025},
      eprint={2502.03874},
      archivePrefix={arXiv},
      primaryClass={quant-ph}}

@article{bong2020strong,
  title={{A strong no-go theorem on the Wigner’s friend paradox}},
  author={Bong, Kok-Wei and Utreras-Alarc{\'o}n, An{\'\i}bal and Ghafari, Farzad and Liang, Yeong-Cherng and Tischler, Nora and Cavalcanti, Eric G and Pryde, Geoff J and Wiseman, Howard M},
  journal={Nat. Phys.},
  volume={16},
  number={12},
  pages={1199--1205},
  year={2020},
  publisher={Nature Publishing Group},
  doi={https://doi.org/10.1038/s41567-020-0990-x}
}

@article{cavalcanti2021implications,
  title={{Implications of Local Friendliness violation for quantum causality}},
  author={Cavalcanti, Eric G and Wiseman, Howard M},
  journal={Entropy},
  volume={23},
  number={8},
  pages={925},
  year={2021},
  publisher={Multidisciplinary Digital Publishing Institute},
  doi={https://doi.org/10.3390/e23080925}
}

@article{brukner2017quantum,
  title={On the quantum measurement problem},
  author={Brukner, {\v{C}}aslav},
  journal={Quantum [Un] Speakables II: Half a Century of Bell's Theorem},
  pages={95--117},
  year={2017},
doi={10.1007/978-3-319-38987-5_5},
  publisher={Springer}
}

@article{nurgalieva2020testing,
  title={Testing quantum theory with thought experiments},
  author={Nurgalieva, Nuriya and Renner, Renato},
  journal={Contemp. Phys.},
  volume={61},
  number={3},
  pages={193--216},
  year={2020},
  publisher={Taylor \& Francis},
  doi={https://doi.org/10.1080/00107514.2021.1880075}
}

@article{vilasini2019multi,
  title={Multi-agent paradoxes beyond quantum theory},
  author={Vilasini, Venkatesh and Nurgalieva, Nuriya and del Rio, L{\'i}dia},
  journal={New J. Phys.},
  volume={21},
  number={11},
  pages={113028},
  year={2019},
  publisher={IOP Publishing},
  doi={10.1088/1367-2630/ab4fc4}
}

@article{nurgalieva2018inadequacy,
	doi = {10.4204/eptcs.287.16},
	year = 2019,
	publisher = {Open Publishing Association},
	volume = {287},
	pages = {267--297},
	author = {Nuriya Nurgalieva and L{\'{\i}}dia del Rio},
	title = {Inadequacy of Modal Logic in Quantum Settings},
	journal = {Electron. Proc. Theor. Comput. Sci.}
}

@article{montanhano2024wigner,
      title={Wigner and friends, a map is not the territory! {C}ontextuality in multi-agent paradoxes}, 
      author={Sidiney B. Montanhano},
      year={2024},
      journal={arXiv:2305.07792},
      url={https://arxiv.org/abs/2305.07792} 
}

@article{allardGuerin2021nogotheorem,
	doi = {10.1038/s42005-021-00589-1},
	url = {https://doi.org/10.1038/s42005-021-00589-1},
	year = 2021,
	month = {may},
	publisher = {Springer Science and Business Media {LLC}},
	volume = {4},
	number = {1},
        pages={93},
	author = {Philippe Allard Gu{\'{e}}rin and Veronika Baumann and Flavio Del Santo and {\v{C}}aslav Brukner},
	title = {{A no-go theorem for the persistent reality of Wigner's friend's perception}},
	journal = {Commun. Phys.}
}

@article{brukner2018no,
  title={A no-go theorem for observer-independent facts},
  author={Brukner, {\v{C}}aslav},
  journal={Entropy},
  volume={20},
  number={5},
  pages={350},
  year={2018},
  publisher={MDPI},
  doi={https://doi.org/10.3390/e20050350}
}

@article{szangolies2020quantum,
  title={The {Q}uantum {R}ashomon Effect: A Strengthened {F}rauchiger-{R}enner Argument},
  author={Szangolies, Jochen},
  journal={arXiv:2011.12716},
  year={2020},
  url={https://arxiv.org/abs/2011.12716}
}

@article{CataniGalleyGonda2024,
      title={Resource-theoretic hierarchy of contextuality for general probabilistic theories}, 
      author={Lorenzo Catani and Thomas D. Galley and Tomáš Gonda},
      year={2024},
      journal={arXiv:2406.00717},
      url={https://arxiv.org/abs/2406.00717} 
}

@article{leegwater2022greenberger,
  title={{When Greenberger, Horne and Zeilinger meet Wigner’s Friend}},
  author={Leegwater, Gijs},
  journal={Found. Phys.},
  volume={52},
  number={4},
  pages={68},
  year={2022},
  publisher={Springer},
  doi={https://doi.org/10.1007/s10701-022-00586-6}
}

@article{utreras2023allowing, title={Allowing {W}igner’s friend to sequentially measure incompatible observables}, volume={480}, ISSN={1471-2946}, url={http://dx.doi.org/10.1098/rspa.2024.0040}, DOI={10.1098/rspa.2024.0040}, number={2288}, pages={(2288) 20240040}, journal={Proc. R. Soc. A}, publisher={The Royal Society}, author={Utreras-Alarcón, Aníbal and Cavalcanti, Eric G. and Wiseman, Howard M.}, year={2024}, 
 month=apr }

@article{walleghem2023extended,
      title={Extended {W}igner's friend paradoxes do not require nonlocal correlations}, 
      author={Laurens Walleghem and Rafael Wagner and Yìlè Yīng and David Schmid},
      year={2024},
      journal={arXiv:2310.06976},
      url={https://doi.org/10.48550/arXiv.2310.06976}
}

@article{walleghem2024connecting,
   title={Connecting extended {W}igner's friend arguments and noncontextuality},
   volume={9},
   ISSN={2521-327X},
   url={http://dx.doi.org/10.22331/q-2025-07-31-1819},
   DOI={10.22331/q-2025-07-31-1819},
   journal={Quantum},
   publisher={Verein zur Forderung des Open Access Publizierens in den Quantenwissenschaften},
   author={Walleghem, Laurens and Yīng, Yìlè and Wagner, Rafael and Schmid, David},
   year={2025},
   month=July, pages={1819} }

@incollection{kochen1990problem,
	doi = {10.1007/978-3-0348-9259-9_21},
	url = {https://doi.org/10.1007/978-3-0348-9259-9_21},
	year = 1990,
	publisher = {Birkhäuser Basel},
	pages = {235--263},
	author = {Simon Kochen and E. P. Specker},
	title = {The Problem of Hidden Variables in Quantum Mechanics},
	booktitle = {Ernst Specker Selecta}
}

@article{brukner2015,
      title={On the quantum measurement problem}, 
      author={Caslav Brukner},
      year={2015},
      journal={arXiv:1507.05255},
      url={https://arxiv.org/abs/1507.05255}, 
}

@article{fine1982hidden,
  title = {Hidden Variables, Joint Probability, and the {B}ell Inequalities},
  author = {Fine, Arthur},
  journal = {Phys. Rev. Lett.},
  volume = {48},
  issue = {5},
  pages = {291--295},
  numpages = {0},
  year = {1982},
  month = {Feb},
  publisher = {American Physical Society},
  doi = {10.1103/PhysRevLett.48.291},
  url = {https://link.aps.org/doi/10.1103/PhysRevLett.48.291}
}

@article{fine1982joint,
  title={Joint distributions, quantum correlations, and commuting observables},
  author={Fine, Arthur},
  journal={J. Math. Phys.},
  volume={23},
  number={7},
  pages={1306--1310},
  year={1982},
  publisher={American Institute of Physics},
  doi = {10.1063/1.525514},
  url={https://doi.org/10.1063/1.525514}
}

@article{ormrod2023theories,
  title={Which theories have a measurement problem?},
  author={Ormrod, Nick and Vilasini, V and Barrett, Jonathan},
  journal={arXiv:2303.03353},
  year={2023},
  url = {https://arxiv.org/abs/2303.03353}
}

@misc{walleghem2026wignersfriendsblackhole,
      title={Wigner's friend's black hole adventure: an argument for complementarity?}, 
      author={Laurens Walleghem},
      year={2026},
      eprint={2507.05369},
      archivePrefix={arXiv},
      primaryClass={gr-qc}}

@inbook{wiseman2017causarum,
  title={Causarum Investigatio and the two {B}ell's theorems of {J}ohn {B}ell},
  author={Wiseman, Howard M and Cavalcanti, Eric G},
editor="Bertlmann, Reinhold
and Zeilinger, Anton",
title="Causarum Investigatio and the Two Bell's Theorems of John Bell",
bookTitle="Quantum [Un]Speakables II: Half a Century of Bell's Theorem",
year="2017",
publisher="Springer International Publishing",
address="Cham",
pages="119--142",
isbn="978-3-319-38987-5",
doi="10.1007/978-3-319-38987-5_6",
url="https://doi.org/10.1007/978-3-319-38987-5_6"
}

@misc{walleghem2026freechoicesabsoluteinternalized,
      title={Are free choices absolute, when internalized in {W}igner's friend?}, 
      author={Laurens Walleghem},
      year={2026},
      eprint={2605.14538},
      archivePrefix={arXiv},
      primaryClass={quant-ph}}

@article{bell1964einstein,
  title={{On the {E}instein {P}odolsky {R}osen paradox}},
  author={Bell, John S},
  journal={Physics Physique Fizika},
  volume={1},
  number={3},
  pages={195},
  year={1964},
doi={10.1142/9789812386540_0002},
  publisher={APS}
}

@article{schmid2023review,
      title={A review and analysis of six extended Wigner's friend arguments}, 
      author={David Schmid and Yìlè Yīng and Matthew Leifer},
      year={2023},
      journal={arXiv:2308.16220},
      url={https://arxiv.org/abs/2308.16220}
}

@article{SchmidSolstice2022,
	title = {Generalized noncontextuality},
	note = {\href{https://www.youtube.com/watch?v=M3qn3EHWdOg}{https://www.youtube.com/watch?v=M3qn3EHWdOg}},
	journal = {Solstice of Foundations, ETH Zurich},
	author = {Schmid, David},
	year = {2022}
}

@article{abramsky2011sheaf,
  title={The sheaf-theoretic structure of non-locality and contextuality},
  author={Abramsky, Samson and Brandenburger, Adam},
  journal={New J. Phys.},
  volume={13},
  number={11},
  pages={113036},
  year={2011},
  publisher={IOP Publishing},
  doi={10.1088/1367-2630/13/11/113036}
}

@article{Chaturvedi_2020,
   title={Quantum prescriptions are more ontologically distinct than they are operationally distinguishable},
   volume={4},
   ISSN={2521-327X},
   url={http://dx.doi.org/10.22331/q-2020-10-21-345},
   DOI={10.22331/q-2020-10-21-345},
   journal={Quantum},
   publisher={Verein zur Forderung des Open Access Publizierens in den Quantenwissenschaften},
   author={Chaturvedi, Anubhav and Saha, Debashis},
   year={2020},
   month=oct, pages={345} }

@article{liang2010specker,
title = {Specker’s parable of the overprotective seer: A road to contextuality, nonlocality and complementarity},
journal = {Physics Reports},
volume = {506},
number = {1},
pages = {1-39},
year = {2011},
issn = {0370-1573},
doi = {https://doi.org/10.1016/j.physrep.2011.05.001},
url = {https://www.sciencedirect.com/science/article/pii/S0370157311001517},
author = {Yeong-Cherng Liang and Robert W. Spekkens and Howard M. Wiseman}
}

@article{wiseman2023thoughtful,
  doi = {10.22331/q-2023-09-14-1112},
  url = {https://doi.org/10.22331/q-2023-09-14-1112},
  title = {A ``thoughtful" {L}ocal {F}riendliness no-go theorem: a prospective experiment with new assumptions to suit},
  author = {Wiseman, Howard M. and Cavalcanti, Eric G. and Rieffel, Eleanor G.},
  journal = {{Quantum}},
  issn = {2521-327X},
  publisher = {{Verein zur F{\"{o}}rderung des Open Access Publizierens in den Quantenwissenschaften}},
  volume = {7},
  pages = {1112},
  month = sep,
  year = {2023}
}

@article{Waaijer2021,
	author = {Waaijer, Marijn and Neerven, Jan van},
	date = {2021/04/03},
	doi = {10.1007/s10701-021-00413-4},
	id = {Waaijer2021},
	isbn = {1572-9516},
	journal = {Found. Phys.},
	number = {2},
	pages = {45},
	title = {Relational Analysis of the {F}rauchiger--{R}enner Paradox and Interaction-Free Detection of Records from the Past},
	url = {https://doi.org/10.1007/s10701-021-00413-4},
	volume = {51},
	year = {2021}}

@article{peres1978unperformed,
  title={Unperformed experiments have no results},
  author={Peres, Asher},
  journal={Am. J. Phys.},
  volume={46},
  number={7},
  pages={745--747},
  year={1978},
  publisher={American Association of Physics Teachers},
url = {https://pubs.aip.org/aapt/ajp/article/46/7/745/1045137/Unperformed-experiments-have-no-results}
}

@Inbook{Wigner1995remarks,
author="Wigner, E. P.",
editor="Mehra, Jagdish",
title="Remarks on the Mind-Body Question",
bookTitle="Philosophical Reflections and Syntheses",
year="1995",
publisher="Springer Berlin Heidelberg",
address="Berlin, Heidelberg",
pages="247--260",
isbn="978-3-642-78374-6",
doi="10.1007/978-3-642-78374-6_20",
url="https://doi.org/10.1007/978-3-642-78374-6_20"
}

@article{Henaut2018,
  title = {Tsirelson's bound and Landauer's principle in a single-system game},
  author = {Henaut, Luciana and Catani, Lorenzo and Browne, Dan E. and Mansfield, Shane and Pappa, Anna},
  journal = {Phys. Rev. A},
  volume = {98},
  issue = {6},
  pages = {060302},
  numpages = {6},
  year = {2018},
  month = {Dec},
  publisher = {American Physical Society},
  doi = {10.1103/PhysRevA.98.060302},
  url = {https://link.aps.org/doi/10.1103/PhysRevA.98.060302}
}

@article{Spekkens2019Leibniz,
      title={The ontological identity of empirical indiscernibles: Leibniz's methodological principle and its significance in the work of {E}instein}, 
      author={Robert W. Spekkens},
      year={2019},
      journal={arXiv:1909.04628},
      url={https://arxiv.org/abs/1909.04628} 
}

@article{CataniFaleiro2023,
  title = {Connecting {XOR} and {XOR}${}^{*}$ games},
  author = {Catani, Lorenzo and Faleiro, Ricardo and Emeriau, Pierre-Emmanuel and Mansfield, Shane and Pappa, Anna},
  journal = {Phys. Rev. A},
  volume = {109},
  issue = {1},
  pages = {012427},
  numpages = {14},
  year = {2024},
  month = {Jan},
  publisher = {American Physical Society},
  doi = {10.1103/PhysRevA.109.012427},
  url = {https://link.aps.org/doi/10.1103/PhysRevA.109.012427}
}

@article{Khoshbin_2024,
title = {Alternative robust ways of witnessing nonclassicality in the simplest scenario},
  author = {Khoshbin, Massy and Catani, Lorenzo and Leifer, Matthew},
  journal = {Phys. Rev. A},
  volume = {109},
  issue = {3},
  pages = {032212},
  numpages = {16},
  year = {2024},
  month = {Mar},
  publisher = {American Physical Society},
  doi = {10.1103/PhysRevA.109.032212},
  url = {https://link.aps.org/doi/10.1103/PhysRevA.109.032212}
}

@article{Wright_2023,
   title = {Invertible Map between {B}ell Nonlocal and Contextuality Scenarios},
  author = {Wright, Victoria J. and Farkas, M\'at\'e},
  journal = {Phys. Rev. Lett.},
  volume = {131},
  issue = {22},
  pages = {220202},
  numpages = {6},
  year = {2023},
  month = {Nov},
  publisher = {American Physical Society},
  doi = {10.1103/PhysRevLett.131.220202},
  url = {https://link.aps.org/doi/10.1103/PhysRevLett.131.220202}
}

@phdthesis{nurgalieva2023multi,
  title={Multi-agent paradoxes in physical theories},
  author={Nuriya Nurgalieva},
  year={2023},
  type = {PhD Thesis},
  school={ETH Zurich},
  url={https://www.research-collection.ethz.ch/handle/20.500.11850/649851?show=full},
  doi={10.3929/ethz-b-000649851}
}

@article{walleghem2024strong,
    title = {A refined {F}rauchiger--{R}enner paradox based on strong contextuality},
    author = {Walleghem, Laurens and Barbosa, Rui Soares and Pusey, Matt and Weigert, Stefan},
    year={2024},
    journal={arXiv:2409.05491},
    url={https://arxiv.org/abs/2409.05491}
}

@article{Maudlin1995,
	author = {Tim Maudlin},
	doi = {10.1007/bf00763473},
	journal = {Topoi},
	number = {1},
	pages = {7--15},
	title = {Three Measurement Problems},
	volume = {14},
	year = {1995}
}

@misc{schmid2021unscrambling,
      title={Unscrambling the omelette of causation and inference: The framework of causal-inferential theories}, 
      author={David Schmid and John H. Selby and Robert W. Spekkens},
      year={2021},
      eprint={2009.03297},
      archivePrefix={arXiv},
      primaryClass={quant-ph}}

@article{harrigan2010einstein,
  title={Einstein, incompleteness, and the epistemic view of quantum states},
  author={Harrigan, Nicholas and Spekkens, Robert W},
  journal={Found. Phys.},
  volume={40},
  pages={125--157},
  year={2010},
  publisher={Springer},
  url = {https://link.springer.com/article/10.1007/s10701-009-9347-0}
}

@book{CoeckeKissinger2017,
title = {Picturing Quantum Processes},
author = {Coecke, Bob and Kissinger, Aleks},
publisher={Cambridge University Press},
year = {2017},
ISBN= {9781316219317},
doi = {https://doi.org/10.1017/9781316219317}
}

@article{catani2023mathematical,
  title={A mathematical framework for operational fine tunings},
   volume={7},
   ISSN={2521-327X},
   url={http://dx.doi.org/10.22331/q-2023-03-16-948},
   DOI={10.22331/q-2023-03-16-948},
   journal={Quantum},
   publisher={Verein zur Forderung des Open Access Publizierens in den Quantenwissenschaften},
   author={Catani, Lorenzo and Leifer, Matthew},
   year={2023},
   month=mar, pages={948} }

@article{pusey2018robust,
  title={Robust preparation noncontextuality inequalities in the simplest scenario},
  author={Pusey, Matthew F},
  journal={Phys. Rev. A},
  volume={98},
  number={2},
  pages={022112},
  year={2018},
  publisher={APS},
   url={http://dx.doi.org/10.1103/PhysRevA.98.022112}
}

@article{pusey2019,
      title={Contextuality without access to a tomographically complete set}, 
      author={Matthew F. Pusey and Lídia del Rio and Bettina Meyer},
      year={2019},
      journal={arXiv:1904.08699},
      url={https://arxiv.org/abs/1904.08699}, 
}

@article{haddara2022possibilistic, 
    title={A possibilistic no-go theorem on the {W}igner’s friend paradox}, 
    volume={25}, 
    url={http://dx.doi.org/10.1088/1367-2630/aceea3}, 
    doi={10.1088/1367-2630/aceea3}, 
    number={9}, 
    journal={New J. Phys.}, 
    publisher={IOP Publishing}, 
    author={Haddara, Marwan and Cavalcanti, Eric G}, 
    year={2023}, 
    month=sep, 
    pages={093028} 
}

@article{spekkens2005contextuality,
  title={Contextuality for preparations, transformations, and unsharp measurements},
  author={Spekkens, Robert W},
  journal={Phys. Rev. A},
  volume={71},
  number={5},
  pages={052108},
  year={2005},
doi={10.1103/physreva.71.052108},
  publisher={APS}
}

@article{haddara2024local,
  title = {Local friendliness polytopes in multipartite scenarios},
  author = {Haddara, Marwan and Cavalcanti, Eric G.},
  journal = {Phys. Rev. A},
  volume = {111},
  issue = {1},
  pages = {012206},
  numpages = {15},
  year = {2025},
  month = {Jan},
  publisher = {American Physical Society},
  doi = {10.1103/PhysRevA.111.012206},
  url = {https://link.aps.org/doi/10.1103/PhysRevA.111.012206}
}

@article{Wood_2015,
   title={The lesson of causal discovery algorithms for quantum correlations: causal explanations of {B}ell-inequality violations require fine-tuning},
   volume={17},
   ISSN={1367-2630},
   url={http://dx.doi.org/10.1088/1367-2630/17/3/033002},
   DOI={10.1088/1367-2630/17/3/033002},
   number={3},
   journal={New J. Phys.},
   publisher={IOP Publishing},
   author={Wood, Christopher J and Spekkens, Robert W},
   year={2015},
   month=mar, pages={033002} }

@article{Cavalcanti_2018,
   title={Classical Causal Models for {B}ell and {K}ochen-{S}pecker Inequality Violations Require Fine-Tuning},
   volume={8},
   ISSN={2160-3308},
   url={http://dx.doi.org/10.1103/PhysRevX.8.021018},
   DOI={10.1103/physrevx.8.021018},
   volume={9},
   pages={021018},
   journal={Phys. Rev. X},
   publisher={American Physical Society (APS)},
   author={Cavalcanti, Eric G.},
   year={2018},
   month=apr }

\appendix
\section{Appendix A: Proof of LF no-go theorem}
\label{AppendixA}

\begin{theorem}[LF no-go theorem] \label{th:LF}
If a superobserver can perform arbitrary quantum operations on an observer and its environment, then no physical theory can satisfy Local Friendliness (the conjunction of AOE and Local Agency).
\end{theorem}
\vspace{-5pt}
\proof{Let us consider the LF scenario of \Cref{fig:LF_cartoon} with the following specifications. Charlie and Debbie share the bipartite entangled state $(|01\rangle-|10\rangle)/\sqrt{2}$, and perform an $(X+Z)/\sqrt{2}$ and a $Z$ measurement, respectively. When $x=1$, Alice undoes Charlie's measurement and performs an $(Z-X)/\sqrt{2}$ measurement; when $y=1$, Bob undoes Debbie's measurement and performs an $X$ measurement (any set of four (sharp) measurements and an entangled state that produce correlations violating a Bell inequality would also work \cite{walleghem2024connecting,haddara2024local}).
There are four empirical distributions:
\begin{equation} \label{eq:empirical_prob_LF}
    \begin{split}
        &\wp(c,d \mid x=0, y=0), \\
&\wp(c,b\mid x=0, y=1), \\
&\wp(a,d \mid x=1, y=0), \\
&\wp(a,b\mid x=1, y=1).
    \end{split}
\end{equation}
By AOE and Local Agency, we can rewrite these
empirical correlations as follows,
     \begin{align}
      \label{eq:empirical_to_global_corr_1}
        &\wp(c,d \mid x=0, y=0) = p(c,d|x=1,y=1), \\  \label{eq:empirical_to_global_corr_2}
&\wp(c,b\mid x=0, y=1)  = p(c,b|x=1,y=1), \\  \label{eq:empirical_to_global_corr_3}
&\wp(a,d \mid x=1, y=0)  = p(a,d|x=1,y=1), \\  \label{eq:empirical_to_global_corr_4}
&\wp(a,b\mid x=1, y=1) = p(a,b|x=1,y=1).
    \end{align}
AOE is used to consider joint probability distributions when $c$ or $d$ are involved and possibly erased. To obtain Eq.~\eqref{eq:empirical_to_global_corr_2} we first appeal to an instance of Local Agency that enforces no-superdeterminism, $p(c|xy) = p(c)$, yielding $p(c|x=0,y=1)=p(c|x=1,y=1)$; we can then apply the rule of conditional probability and consider another instance of Local Agency expressing that neither $b$ nor $c$ occur in the future light cone of $x$: $p(b|c,x,y){=}p(b|c,y)$,
thus obtaining \[\begin{split}&\wp(c,b|x{=}0,y{=}1) = p(c|x{=}0,y{=}1)p(b|c,x{=}0,y{=}1) \\&=p(c|x{=}1,y{=}1)p(b|c,x{=}1,y{=}1)=p(c,b|x{=}1,y{=}1).\end{split}\]

Similar instances of Local Agency lead to Eq.~\eqref{eq:empirical_to_global_corr_3}.
Eq.~\eqref{eq:empirical_to_global_corr_1} derives from an instance of Local Agency enforcing no-superdeterminism, $p(c,d|x,y)=p(c,d)$.

By assuming AOE, we can assert that all the correlations in Eqs.~\eqref{eq:empirical_to_global_corr_1}, \eqref{eq:empirical_to_global_corr_2}, \eqref{eq:empirical_to_global_corr_3} and \eqref{eq:empirical_to_global_corr_4} can be obtained
from a single joint probability distribution $p(a, b, c, d|x =
1, y = 1)$ via marginalization.
However, given the choices of preparations and measurements made, by Fine's theorem~\cite{fine1982hidden,abramsky2011sheaf}, such distribution cannot exist.
Indeed, these give exactly the empirical correlations that lead to a maximal violation of the Bell inequalities.
Let us stress again, though, that the LF assumptions are weaker than those of Bell's theorem. Only the empirical correlations in the scenarios are the same.}

\section{Appendix B: Noncontextual Friendliness imposes weaker constraints on correlations than noncontextuality}
\label{AppendixB}

We now prove that the assumptions entering the Noncontextual Friendliness no-go theorem are strictly weaker than the set of assumptions for noncontextuality inequalities. Thus, granted that NF scenarios can be realized, the NF no-go theorem leads to stronger conclusions than no-go theorems relying on noncontextuality. The proof works by constructing a scenario where correlations are consistent with AOE and Noncontextual Agency but not with the existence of a noncontextual ontological model. This scenario extends the NF setup from \Cref{fig:OF_tryout} by introducing additional measurement choices for Alice and Bob. Our strategy mirrors that of \cite{haddara2024local}, which demonstrated that the LF correlations form a polytope that strictly contains the polytope of local correlations.

Consider the NF setup of \Cref{fig:OF_tryout}. In addition to Alice's existing choices associated with $x=0,1$ -- corresponding to asking for Charlie's outcome and performing $U_C^\dagger$, respectively -- we introduce a third option, associated with $x=2$, allowing her to perform another operation (which may include undoing $U_C$). Similarly, besides Bob's choices  associated with $y=0,1$ -- corresponding to asking for Debbie's outcome and performing $U_D^\dagger$ followed by a measurement, respectively -- we introduce a third option, associated with $y=2$, allowing him to perform another operation.
Recall that $p(a|x=0) = p(c|x=0)$ and $p(b|a,x,y=0)=p(d|a,x,y=0)$.

\begin{figure*}[t]
         \centering \includegraphics[width=0.8\textwidth]{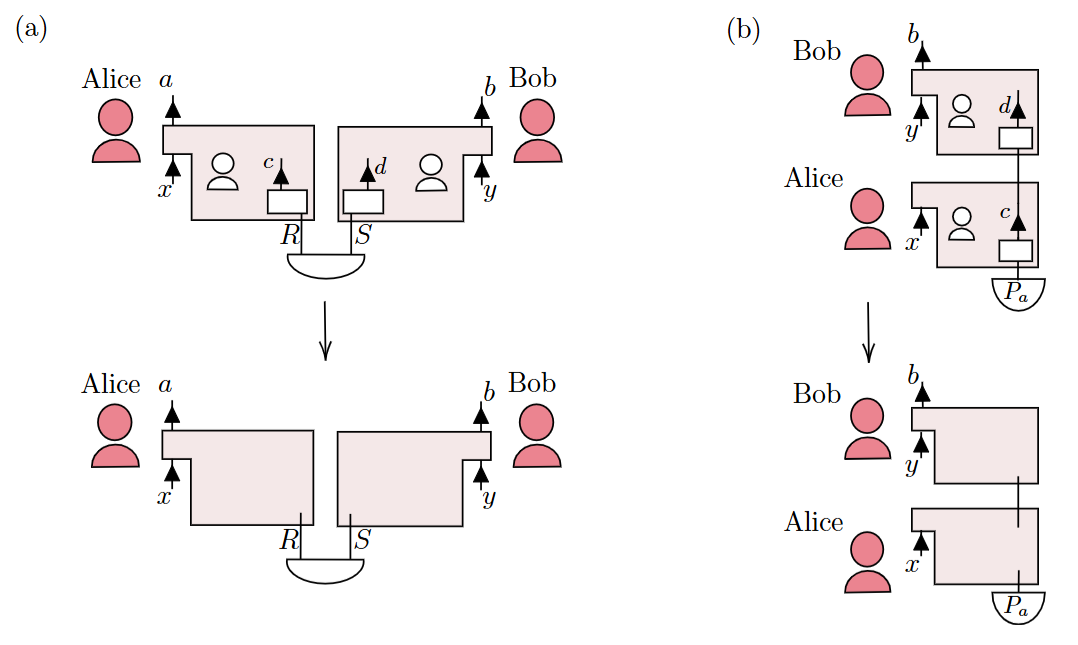}
         \vspace{0.2cm}
         \caption{(a): A LF scenario (top) can be seen as a Bell scenario (bottom) where Charlie's and Debbie's operations are part of Alice's preparation and Bob's measurement, respectively. (b): Similarly, an NF scenario (top) can be seen as a prepare-and-measure scenario (bottom) where Charlie's and Debbie's operations are part of Alice's preparation and Bob's measurement, respectively.
         }
         \label{fig:LF_to_Bell_OF_to_PM}
\end{figure*}

The empirical (verifiable) correlations that satisfy the operational equivalences relevant for the NF no-go theorem involving $b,y$ and $x$ are \begin{equation}
\begin{split}
    \wp(b|y=1,x=1) &= \wp(b|y=1,x=0), \\ \wp(d|y=0,x=1) &= \wp(d|y=0, x=0).
\end{split}
\end{equation}
As part of the setup, we further assume that, for $y=2$,
\begin{equation}
\label{eq:NewOE}
    \wp(b|y=2,x=1) = \wp(b|y=2,x=0).
\end{equation}
To satisfy equation \eqref{eq:NewOE}, it suffices that the preparations associated with the two values of $a$ and the ones associated with the two values of $c$ mix into the same mixed state.
Additionally, Noncontextual Agency requires: \begin{equation}
\begin{split}
    p(b|y=1,c,x=1) &= p(b|y=1,c,x=0), \\  p(b|y=2,c,x=1) &= p(b|y=2,c,x=0) \\  p(d|c,x=1) &= p(d|c,x=0).
\end{split}
\end{equation}

In fact, these equalities are predicted by the theory for any alternative protocol in which $c$ is recorded in a separate register prior to Alice's operation, since in that case Alice's actions do not affect the recorded outcome. By Noncontextual Agency, they continue to hold for the empirical-level correlations in the actual protocol, where these joint variables are inaccessible to any single agent but, by virtue of AOE, still exist as facts about the world; in particular, they hold regardless of whether Bob chooses $y=1$ or $y=2$.
However, notice that when Alice performs the operation corresponding to $x=2$, Noncontextual Friendliness does not prevent the possibility that, in general, the theory itself would predict
\begin{equation}
    p(d|c,x=0) \neq p(d|c,x=2) \neq p(d|c,x=1),
\end{equation} and similarly for $b,c$.
Let us see this in a specific example. Consider the LF setup from the proof of Theorem \ref{th:NCF} and imagine that Alice, for $x=2,$ performs a rotation $R_\varphi$ by an angle $\varphi  \neq 0$ around the $y$-axis on the system measured by Charlie. The theory then predicts that $p(d|c,x=1)$ is obtained by performing the measurement $D$ on $U_C^\dagger \ket{c}$ -- which yields the same result as performing a measurement on $\ket{c}$ -- whereas $p(d|c,x=2)$ is obtained by performing the measurement $D$ on $R_\varphi \ket{c}$. These two distributions are in general distinct, so $p(d|c,x=1) \neq p(d|c,x=2)$. An analogous argument works when considering $b,c$.

In conclusion, Noncontextual Friendliness poses no restrictions (apart from no-superdeterminism) on the correlations $p(b,a|y=1,x=1),p(b,a|y=2,x=1),p(b,a|y=1,x=2),p(b,a|y=2,x=2)$. Therefore, these correlations can violate noncontextuality inequalities while the scenario still satisfies Noncontextual Friendliness.

\begin{figure}[h]
         \centering \includegraphics[width=0.45\textwidth]{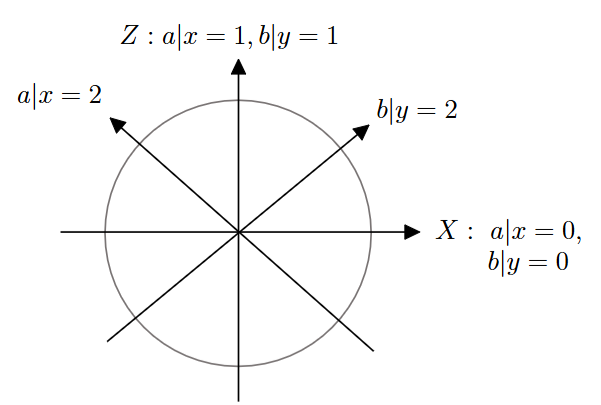}
         \caption{ The figure represents the preparations and measurements involved in the scenario showing how NF holds while noncontextuality is violated.
         Specifically, Alice's outcomes $a=0,1$ given her operations associated with $x$, $a|x=0$ and $a|x=1$, correspond to the eigenvectors of the $X$ and $Z$ measurements, respectively, and $a|x=2$ to the eigenvectors of the $Z-X$ measurement. Bob's outcomes $b=0,1$ given his operations associated with $y$, $b|y=0$ and $b|y=1$, correspond to the $X$ and $Z$ measurements, and $b|y=2$ to the $X+Z$ measurement.
         }
         \label{fig:example_app}
\end{figure}

Let us show this result with a specific example (see \Cref{fig:example_app}). Let the system $S$ be prepared in the state $P_a$, the state inputted to Charlie's lab, corresponding to one of the Pauli $Z$ eigenstates, and let Charlie perform a measurement in the $X$ basis. For $x=0$, Alice takes Charlie's outcome as her own, $a=c$, while, for $x=1$, Alice undoes Charlie's measurement. For $x=2$, Alice undoes Charlie's outcome and applies the unitary that performs a rotation of $\pi/4$ around the $y$-axis. Subsequently, the system $S$ is passed to Debbie who performs a measurement on the $X$ basis. For $y=0$, Bob takes Debbie's outcome as his own, $b=d$, while, for $y=1$, Bob undoes Debbie's measurement and performs a measurement in the $Z$ basis. For $y=2$, Bob undoes Debbie's measurement and performs a measurement in the $X+Z$ basis.
The scenario just described does not admit of a noncontextual ontological model because the empirical correlations associated with $x=1,x=2,y=1,y=2$ are those of the simplest nontrivial scenario that provide a maximal violation of the noncontextuality inequalities \cite{pusey2018robust}.
However, Noncontextual Friendliness remains satisfied. Specifically, it is consistent with
\begin{equation}
\begin{split}
        &p(d|c,x{=}1,y{=}0) \neq p(d|c,x{=}2,y{=}0) \neq p(d|c,x{=}0,y{=}0), \\
        &p(b|c,x{=}1,y{=}1) \neq p(b|c,x{=}2,y{=}1) \neq p(b|c,x{=}0,y{=}1).
\end{split}
\end{equation}

This means that there are no operational equivalences for which Noncontextual Agency would require the existence of a global distribution reproducing the correlations $\wp(b,c|x,y)$ for $x,y \in \{1,2\}$.
Instead, Noncontextual Agency \textit{does} require the existence of a global distribution $p(a,b,c,d|x=1,y=1)$ that reproduces the empirical correlations $\wp(a,b|x=1,y=1),\wp(c,b|x=0,y=1),\wp(a,d|x=1,y=0),\wp(c,d|x=0,y=0)$.
Such a global distribution indeed exists, as these correlations are simply the ones associated with $X,Z$ eigenstates and $X,Z$ measurements.

Noncontextual Agency additionally requires no-superdeterminism, but the only new constraints imposed by no-superdeterminism in the extended setup including $x=2,y=2$ are \begin{equation}
\begin{split}
       p(a,c,d|x=2,y) &= p(a,c,d|x=2), \\ p(a,c,d|x,y=2)
&= p(a,c,d|x), \\ p(a,c|x=2) &= p(a,c|x=1).
\end{split}
\end{equation} All no-superdeterminism requirements can be satisfied by constructing a distribution $P$ such that $P(a,c,d|x=2,y) = \wp(d|x=2,y=0,a) p(a,c|x=1,y=1)$, with $p(a,c|x=1,y=1)$ coming from the global distribution $p(a,b,c,d|x=1,y=1)$, $P(a,c,d|x=1,y) = p(a,c,d|x=1,y=1)$ and $P(a,c,d|x=0,y)=p(a,c,d|x=1,y=1)$.
Therefore, by this definition we have \begin{equation}
    \begin{split}
        P(a,c,d|x{=}2,y) &= \wp(d|x{=}2,y{=}0,a) p(a,c|x{=}1,y{=}1) \\ &= P(a,c,d|x{=}2,y'), \\
        P(a,c,d|x{=}1,y{=}2) &= p(a,c,d|x{=}1,y{=}1) \\ &= P(a,c,d|x{=}1,y{=}2) \\
        P(a,c|x{=}2) &= p(a,c|x{=}1,y{=}1) = P(a,c|x{=}1).
    \end{split}
\end{equation}
In this scenario, all that Noncontextual Friendliness requires is the existence of a global distribution $p(a,b,c,d|x=1,y=1)$ that reproduces the empirical correlations $\wp(a,b|x=1,y=1),\wp(a,d|x=1,y=0),\wp(c,b|x=0,y=1),\wp(c,d|x=0,y=0)$ and no-superdeterminism, but it does \emph{not} require the existence of one single global distribution that reproduces all the empirical correlations $\wp(a,b|x=1,y=1),\wp(a,d|x=1,y=0),\wp(c,b|x=0,y=1),\wp(c,d|x=0,y=0),\wp(a,b|x=2,y=2),\wp(a,d|x=2,y=0),\wp(c,b|x=0,y=2),\wp(c,d|x=0,y=0)$.

An alternative protocol to the one we just provided would be to keep Alice's choice variable $x$ to range only in $\{0,1\}$, and then for Bob, $y=0,1$, but to include an extra choice variable $z=0,1$ that allows Alice to choose from additional preparations. Then, similarly to the argument above, NF would impose no requirement on the empirical correlations other than no-superdeterminism, $\wp(a,b|x=1,z=1,y=1), \wp(a,b|x=1,z=0,y=1), \wp(a,d|x=1,z=1,y=0), \wp(a,d|x=1,z=0,y=0)$. Thus, the correlations $\wp(a,b|x=1,z=1,y=1), \wp(a,b|x=1,z=0,y=1), \wp(a,d|x=1,z=1,y=0), \wp(a,d|x=1,z=0,y=0)$ can violate noncontextuality inequalities while satisfying Noncontextual Friendliness. The latter indeed requires the existence of global distributions $p(a,b,c,d|x=1,y=1)$ and $p(a,b,c,d|x=2,y=2)$ that reproduce the empirical correlations $\wp(a,b|x=1,y=1,z=0),\wp(a,d|x=1,y=0,z=0),\wp(c,b|x=0,y=1,z=0),\wp(c,d|x=0,y=0,z=0)$ and $\wp(a,b|x=1,y=1,z=1),\wp(a,d|x=1,y=0,z=1),\wp(c,b|x=0,y=1,z=1),\wp(c,d|x=0,y=0,z=1)$, but it does \emph{not} require the existence of one single global distribution that reproduces all these empirical correlations.

We conclude by referring to \Cref{fig:LF_to_Bell_OF_to_PM}, which illustrates how an LF scenario can be viewed as a Bell scenario involving more complex measurement settings for Alice and Bob (\Cref{fig:LF_to_Bell_OF_to_PM}(a)) and how an NF scenario can be viewed as a prepare-and-measure scenario involving more complex measurement settings for Alice and Bob (\Cref{fig:LF_to_Bell_OF_to_PM}(b)). In \cite{bong2020strong} it is shown that the existence of a locally causal ontological model reproducing the empirical distributions $p(a,b|x,y)$ constitutes a stronger requirement than the Local Friendliness assumption \textit{when Alice and Bob have more than two settings choices} $x,y \in \{1,2,3,\ldots\}$. In this appendix, we have demonstrated that the same holds with respect to the existence of a noncontextual ontological model and the Noncontextual Friendliness assumptions.

\end{document}